\documentclass{article}
\usepackage{amssymb}
\usepackage{amsmath}
\usepackage{graphicx}

\begin{document}

\title{Long range dependence and the dynamics of exploited fish populations}
\author{Hugo C. Mendes\thanks{%
Instituto Portugu\^{e}s do Mar e da Atmosfera, Avenida Bras\'{\i}lia,
1300-598 Lisboa} \thanks{%
hugart@gmail.com}, Alberto Murta\footnotemark[1] and R. Vilela Mendes\thanks{%
CMAF and IPFN, Univ. Lisbon, Av. Prof. Gama Pinto 2, 1649-003 Lisboa} 
\thanks{%
rvmendes@fc.ul.pt}}
\date{ }
\maketitle

\begin{abstract}
Long range dependence or long memory is a feature of many processes in the
natural world, which provides important insights on the underlying
mechanisms that generate the observed data. The usual tools available to
characterize the phenomenon are mostly based on second order correlations.
However, the long memory effects may not be evident at the level of second
order correlations and may require a deeper analysis of the nature of the
stochastic processes.

After a short review of the notions and tools used to characterize long
range dependence, we analyse data related to the abundance of exploited fish
populations which provides an example of higher order long range dependence.
\end{abstract}

\section{Introduction}

Long range dependence or long memory is an important notion in many
processes in the natural world. Studies involving this notion pervade fields
from biology to econometrics, linguistics, hydrology, climate, DNA
sequencing, etc. Although, at times, it has been considered a nuisance in
the study of these processes, the existence of long memory is in fact a
bonus, in the sense that it provides further insight in the nature of the
process. Whereas short or no memory just points out to the essentially
unstructured nature of the phenomena, long memory, by contrast, may provide
a window on the underlying mechanisms that generate the observed data.

The most popular definitions of long range dependence are based on the
second-order properties of the processes and relate to the asymptotic
behavior of covariances, spectral densities and variances of partial sums.
However there are other different points of view, some of which are not
equivalent to the characterization of second-order properties. They include
ergodic theory notions, limiting behavior, large deviations, fractional
differentiation, etc. \cite{Samorod} \cite{Dominique} \cite{TaqquBook}.

When looking for or extracting long range dependence from a time series, two
important warnings should be taken into account. First, long range
dependence may be mimicked by lack of stationarity or by a change of regime.
Long range dependence is a notion that refers to stationary processes.
Second, long range dependence is a feature which may only be associated to
the higher order characteristics of the process. A process that looks short
range when looked at through second-order properties, may in fact have an
underlying long range dependence of higher order properties.

In Section 2 we collect a few definitions of the most usual parameters, used
to characterize long range dependence, that will be useful later on and in
Section 3 analyze some data related to the abundance of exploited fish
populations which provides an example of higher order long range dependence.

\section{Notions and tools for long-range dependence}

When based on second order properties, long range dependence in a stationary
time series $X\left( t\right) $ occurs when the covariances%
\begin{equation}
\gamma \left( \tau \right) =\mathbb{E}\left\{ X\left( t\right) X\left(
t+\tau \right) \right\}  \label{2.1}
\end{equation}%
tend to zero so slowly that their sum%
\begin{equation}
\sum_{\tau =0}^{\infty }\gamma \left( \tau \right)  \label{2.2}
\end{equation}%
diverges. Alternative definitions of long-range dependence are based on the
power-law behavior of the covariances, namely%
\begin{equation}
\begin{array}{cccccc}
\sum_{\tau =-n}^{n}\gamma \left( \tau \right) & \backsim & n^{\alpha
}L_{1}\left( n\right) & \text{when} & n\rightarrow \infty & ;0<\alpha <1 \\ 
\gamma \left( \tau \right) & \backsim & \tau ^{-\beta }L_{2}\left( \tau
\right) & \text{when} & \tau \rightarrow \infty & ;0<\beta <1 \\ 
f\left( \nu \right) & \backsim & \left\vert \nu \right\vert ^{-\xi
}L_{3}\left( \left\vert \nu \right\vert \right) & \text{when} & \nu
\rightarrow 0 & ;0<\xi <1%
\end{array}
\label{2.3}
\end{equation}%
$f\left( \nu \right) $ being the spectral density%
\begin{equation}
f\left( \nu \right) =\frac{1}{2\pi }\sum_{\tau =-\infty }^{\infty }e^{-i\nu
\tau }\gamma \left( \tau \right)  \label{2.4}
\end{equation}%
$L_{1},L_{2}$ being slowly varying functions at infinity and $L_{3}$ slowly
varying at zero.

If $\gamma \left( \tau \right) $ is monotone as $\tau \rightarrow \infty $,
the definitions (\ref{2.3}) are equivalent to the divergence of the sum in (%
\ref{2.2}) with $\alpha =1-\beta $ and $\xi =1-\beta $.

In science, one of the main purposes when observing natural phenomena is the
construction of models. A useful approach in this endeavour is the
comparison of natural time series with the behavior of well-studied
mathematical structures. In the context of long range dependence a central
role is played by the theory of self-similar stochastic processes%
\begin{equation}
X\left( at\right) \overset{d}{=}a^{H}X\left( t\right)  \label{2.5}
\end{equation}%
with stationary increments%
\begin{equation}
X\left( t+h\right) -X\left( t\right) \overset{d}{=}X\left( t\right) -X\left(
0\right)  \label{2.6}
\end{equation}%
$\overset{d}{=}$ meaning equality in distribution. These processes are
denoted as $H-sssi$ processes and $H$ is called the Hurst exponent. Notice
that there is a close relation between self-similarity and stationarity. If $%
X\left( t\right) $ is self-similar then $Y\left( t\right) =e^{-tH}X\left(
t\right) $ is stationary and conversely if $Y\left( t\right) $ is stationary 
$X\left( t\right) =t^{H}Y\left( \ln t\right) $ is self-similar.

A finite $\sigma ^{2}$ variance $H-sssi$ process has a covariance%
\begin{equation}
\mathbb{E}\left\{ X\left( s\right) X\left( t\right) \right\} =\frac{\sigma
^{2}}{2}\left\{ \left\vert t\right\vert ^{2H}+\left\vert s\right\vert
^{2H}-\left\vert t-s\right\vert ^{2H}\right\}   \label{2.7}
\end{equation}%
Throughout this paper $\mathbb{E}\left\{ \cdots \right\} $ will denote the
expected value which, for all the data examples, one approximates by the
empirical average $\left\langle \cdots \right\rangle $.

There are non-Gaussian $H-sssi$ processes as well as $H-sssi$ processes with
infinite covariance \cite{SamoTaqqu}. However, the simplest example of a $%
H-sssi$ process is a Gaussian process uniquely defined by the covariance (%
\ref{2.7}) and normalized to have $\sigma ^{2}=1$. It is called \textit{%
fractional Brownian motion} (fBm) $B_{H}\left( t\right) $ and the increment
process%
\begin{equation}
Z\left( t\right) =B_{H}\left( t+1\right) -B_{H}\left( t\right)   \label{2.8}
\end{equation}%
is called \textit{fractional Gaussian noise} (fGn). For $H=\frac{1}{2}$, $%
B_{1/2}\left( t\right) $ is Brownian motion. Fractional Gaussian noise has
covariance%
\begin{equation}
\gamma \left( \tau \right) =\frac{1}{2}\left\{ \left\vert \tau +1\right\vert
^{2H}-2\left\vert \tau \right\vert ^{2H}+\left\vert \tau -1\right\vert
^{2H}\right\}   \label{2.9}
\end{equation}%
hence, if $H=\frac{1}{2}$ it has $\gamma \left( \tau \right) =0$ (no memory)
and for $H\neq \frac{1}{2}$ and large $\tau $%
\begin{equation}
\gamma \left( \tau \right) \backsim H\left( 2H-1\right) \left\vert \tau
\right\vert ^{2H-2}\hspace{1.5cm}\tau \rightarrow \infty   \label{2.10}
\end{equation}%
For $\frac{1}{2}<H<1$ the process has long range dependence. Because $\gamma
\left( \tau \right) >0$ for $H>\frac{1}{2}$ and $\gamma \left( \tau \right)
<0$ for $H<\frac{1}{2}$, the process is called \textit{persistent} in the
first case and \textit{anti-persistent} in the second. For the spectral
function at $H\neq \frac{1}{2}$%
\begin{equation}
f\left( \nu \right) \backsim \nu ^{1-2H}\hspace{1.5cm}\nu \rightarrow 0
\label{2.11}
\end{equation}%
which for $H>\frac{1}{2}$ blows up at the origin.

The Hurst exponent ($H$) as an index of long range dependence quantifies the
tendency of a time series either to regress strongly to the mean or to
persist in a deviation from the mean. An $H$ value between $0.5$ and $1$
signals a time series with long-term positive autocorrelation, meaning that
a high (or low) value in the series will probably be followed by another
high (or low) value. A value in the range $0$ to $0.5$ signals long-term
switching between high and low values, meaning that an high value will
probably be followed by a low value, with this tendency to switch between
high and low values lasting into the future.

Although the correlation behavior of the fractional processes is very
different from simple Brownian motion, they may be represented as integrals
of Brownian motion with the appropriate integration kernel. For example%
\begin{eqnarray}
B_{H}\left( t\right) &=&\int_{-\infty }^{0}\left\{ \left( t-u\right) ^{H-%
\frac{1}{2}}-\left( -u\right) ^{H-\frac{1}{2}}\right\} dB\left( u\right) 
\notag \\
&&+\int_{0}^{t}\left( t-u\right) ^{H-\frac{1}{2}}dB\left( u\right)
\label{2.12}
\end{eqnarray}%
where $B\left( u\right) =B_{\frac{1}{2}}\left( u\right) $. One practical
implication is that to extract an eventual fractional behavior from the data
it is not sufficient an observation of short time intervals where the
process may easily be confused with an uncorrelated process.

Another way in which Brownian motion intervenes in modelling processes which
are neither uncorrelated nor simple Brownian motion is through the following
fundamental result of stochastic analysis \cite{Nualart}: If $X\left(
t\right) $ is a random variable that is square-integrable in the measure
generated by Brownian motion, then%
\begin{equation}
dX\left( t\right) =\mu \left( t\right) dt+\sigma \left( t\right) dB\left(
t\right)  \label{2.13}
\end{equation}%
where $\mu \left( t\right) $ and $\sigma \left( t\right) $ are well defined
stochastic processes. Therefore although the increments of $X\left( t\right) 
$ have a representation in terms of the increments of Brownian motion, the
process may be very different, depending on the nature of the processes $\mu
\left( t\right) $ and $\sigma \left( t\right) $.

Whenever long range dependence is modelled by fractional Gaussian noise one
benefits from the extensive theoretical and computational framework that is
available for this process. However, fractional Gaussian noise is quite
rigid in the sense that it specifies the correlations at all time lags, not
only at $\tau \rightarrow \infty $. It may therefore not be suitable for
modeling long range dependent phenomena where the covariance at short time
lags differs from fGn. This motivated the development of other models
through \textit{Gaussian linear sequences}%
\begin{equation}
X\left( t\right) =\sum_{j=-\infty }^{\infty }c_{t-j}\epsilon _{j}
\label{2.14}
\end{equation}%
where $\sum_{j=-\infty }^{\infty }c_{j}^{2}<\infty $ and $\left\{ \epsilon
_{j}\right\} _{j\in \mathbb{Z}}$ are independent identically distributed
(i.i.d.) normal random variables, called \textit{innovations}. A Gaussian
linear sequence is stationary and for the convergence of the sum in (\ref%
{2.14}) one requires

- If $\epsilon _{j}=N\left( \mu ,\sigma ^{2}\right) $ with $\mu \neq 0$, $%
\sum_{j}\left\vert c_{j}\right\vert <\infty $

- If $\epsilon _{j}=N\left( \mu ,\sigma ^{2}\right) $ with $\mu =0$, $%
\sum_{j}\left\vert c_{j}\right\vert ^{2}<\infty $

An example is the FARIMA(p,d,q) process (fractional autoregressive
integrated moving average) \cite{Granger} \cite{Hosking}%
\begin{equation}
X\left( t\right) =\Phi _{p}^{-1}\left( S\right) \Theta _{q}\left( S\right)
\Delta ^{-d}\epsilon _{t}\hspace{1.5cm}t\in \mathbb{Z}  \label{2.15}
\end{equation}%
$\left\{ \epsilon _{t}\right\} $ is an i.i.d. $N\left( 0,\sigma ^{2}\right) $
sequence and $\Phi _{p}\left( S\right) ,\Theta _{q}\left( S\right) $ are
polynomials on the shift operator%
\begin{equation}
S\epsilon _{i}=\epsilon _{i-1}  \label{2.16}
\end{equation}%
$\Delta ^{-d}$ being%
\begin{equation}
\Delta ^{-d}=\left( \mathbf{1-S}\right) ^{-d}=\sum_{i=0}^{\infty }\frac{%
\Gamma \left( i+d\right) }{\Gamma \left( d\right) \Gamma \left( i+1\right) }%
S^{i}  \label{2.17}
\end{equation}%
The fractional differencing $\Delta ^{-d}$ for $0<d<\frac{1}{2}$ models long
range dependence, whereas the auto regressive $\Phi _{p}\left( S\right) $
and the moving average $\Theta _{q}\left( S\right) $ polynomials provide
flexibility in modeling the short-range dependence.

Finally, as mentioned on the introduction, there are other ways to deal with
long range dependence for which the behavior of covariances does not play
the main role. A potentially promising way is based on ergodic theory
because the notion of memory is related to the connection between a process
and its shifts. Then, a possible definition of long range dependent process
would be one that is ergodic but non-mixing. However the mixing property is
probably not sufficiently strong to imply that a mixing stationary process
has short memory. Stronger requirements may be needed. These notions will
not be used here and we refer to \cite{Samorod} \cite{Bradley} for a
discussion.

\section{The dynamics of exploited fish populations}

Long range dependence has been rarely documented in marine ecology,
presumably because of the scarcity of long time series. This lack of
extended time series has limited research on long memory in fish stock
sizes, whose fluctuations are more often attributed to human exploitation,
because most studies focus on highly exploited populations (such as the
North Atlantic stocks) and over relatively short time periods. However, for
a few fish populations, studies on long-term fluctuations have found long
ranging trends related to human activity, mostly through overexploitation
and pollution of spawning and nursery areas, environmental changes that
affect the recruitment period inducing natural fluctuations in stock size
and biotic processes, such as predation, cannibalism and competition \cite%
{Dickson} \cite{Bjornstad} \cite{Fromentin}.

A few years ago Niwa \cite{Niwa} studying the time series of 27 commercial
fish stocks in the North Atlantic concluded that the variability in the
population growth (the annual changes in the logarithm of population
abundance $S\left( t\right) $)%
\begin{equation}
r\left( t\right) =\ln \left( \frac{S\left( t+1\right) }{S\left( t\right) }%
\right)   \label{3.1}
\end{equation}%
is described by a Gaussian distribution. That is, the population variability
process would be a geometric random walk%
\begin{equation}
r\left( t\right) =\frac{dS\left( t\right) }{S\left( t\right) }=\sigma
_{r}dB\left( t\right)   \label{3.2}
\end{equation}%
for some constant $\sigma _{r}$ depending on the species. The independence
of the increments of Brownian motion would then imply that $r\left( t\right) 
$ is a purely random process.

If completely accurate this would be a sobering conclusion. Natural
processes that look purely random, are processes that depend on some many
uncontrollable variables that any attempt to handle them is outside our
reach. This would be a serious blow to, for example, the implementation of
sustentability measures.

In this paper we reanalyze some of the same type of data to confirm or
sharpen the conclusions in \cite{Niwa}. To explore the variability in fish
population growth we extracted information on the Spawning-stock biomass
(SSB) data on commercial fish stocks in the North Atlantic. The available
SSB time-series data are derived from age-based analytical assessments
estimated by the 2013 working groups of the International Council for the
Exploration of the Sea (ICES), based on the compilation of relevant data
from sampling of fisheries (e.g. commercial catch-at-age) and from
scientific research surveys. From the collection of available assessment
data we selected three North Atlantic stocks for which the annual
time-series of SSB covers at least 60 years, namely Northeast Arctic cod
(Gadus morhua), Arctic haddock (Melanogrammus aeglefinus) and the North Sea
autumn-spawning herring (Clupea harengus). At present, ICES classifies these
stocks as having above average biomass levels with full reproductive
capacity and being harvested sustainably under active management plans. The
stock assessment detailed information is available at the ICES webpage \cite%
{ICES}.

For these three species we analyze the autocorrelation functions for $%
r\left( t\right) $ and $\left\vert r\left( t\right) \right\vert $%
\begin{equation}
C\left( r,\tau \right) =\frac{\mathbb{E}\left\{ r\left( t\right) r\left(
t+\tau \right) \right\} }{\sigma ^{2}}  \label{3.3}
\end{equation}%
\begin{equation}
C\left( \left\vert r\right\vert ,\tau \right) =\frac{\mathbb{E}\left\{
\left\vert r\left( t\right) \right\vert \left\vert r\left( t+\tau \right)
\right\vert \right\} }{\sigma ^{2}}  \label{3.4}
\end{equation}%
The results are shown in the Fig.1

\begin{figure}[p]
\centering
\includegraphics[width=0.8\textwidth]{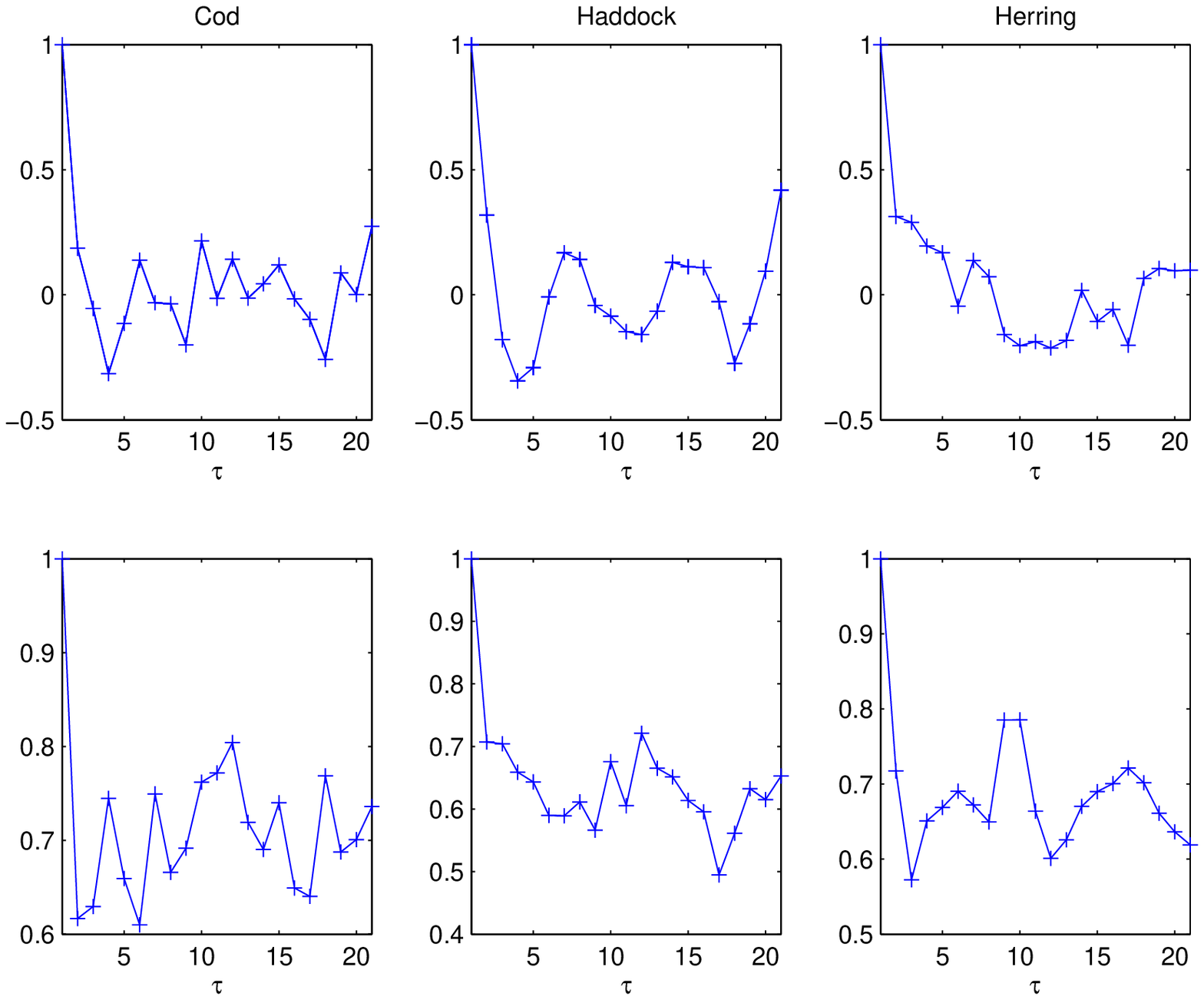}
\caption{Autocorrelation functions for $r\left( t\right) $ (upper plot) and $%
\left\vert r\left( t\right) \right\vert $ (lower plot)}
\end{figure}

One sees that, already for time lags of one year, autocorrelations are at
noise level, suggestive of uncorrelated processes. However, if $S\left(
t\right) $ is indeed a geometrical Brownian motion, to rely on correlations
or fitting of probability distribution functions is not sufficient. The
scaling properties of $r\left( t\right) $ should be checked. As Niwa \cite%
{Niwa} rightly points out, defining%
\begin{equation}
r_{\Delta }\left( t\right) =\ln \left\{ \frac{S\left( t+\Delta \right) }{%
S\left( t\right) }\right\} =\sum_{i=1}^{\Delta }r\left( t+i\right) 
\label{3.5}
\end{equation}%
the geometrical Brownian motion hypothesis would imply%
\begin{equation}
\left( \mathbb{E}\left\{ r_{\Delta }^{2}\right\} \right) ^{1/2}\backsim
\Delta ^{1/2}  \label{3.6}
\end{equation}%
Normalizing $\mathbb{E}\left\{ r_{\Delta }^{2}\right\} $ by the covariance $%
\sigma _{r}^{2}$ for each species and taking the average over all species,
Niwa has obtained a behavior roughly consistent with (\ref{3.6}). However,
when we analyzed each species separately, the hypothesis does not hold. In
Fig.2 we have plotted in loglog scale the computed $\left( \mathbb{E}\left\{
r_{\Delta }^{2}\right\} \right) ^{1/2}$ as a function of $\Delta $ for the
three species analyzed in this paper.

\begin{figure}[p]
\centering
\includegraphics[width=0.8\textwidth]{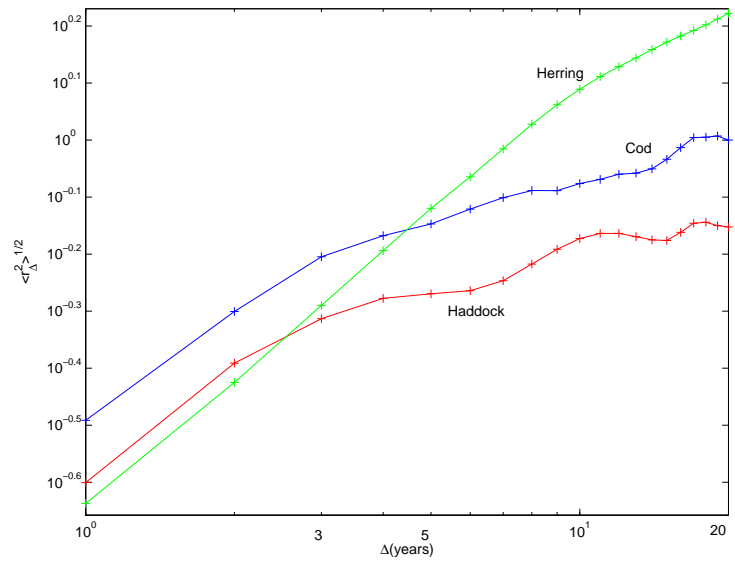}
\caption{$\left( \mathbb{E}\left\{ r_{\Delta }^{2}\right\} \right) ^{1/2}$
as a function of $\Delta $}
\end{figure}

One sees that at the species level the geometrical Brownian motion is not a
good hypothesis. Even for Herring, where the data seems to follow a scaling
law, the slope at large $\Delta $ is closer to $0.7$ than to $0.5$. The
conclusion is that whatever is actually determining the stochastic process
for each species is somehow washed out when averaging over all the 27
species as Niwa did.

Actually this is no surprise. Recall the stochastic analysis result (\ref%
{2.13}). To find a process $r\left( t\right) $ that has features close to
Brownian motion, but is not exactly Brownian motion only means that the
process is square-integrable with respect to the (Wiener) measure generated
by Brownian motion. All the interesting features actually lie on the
dynamics of the process $\sigma \left( t\right) $. That is, on the dynamics
of the amplitude of the fluctuations. In fact this makes sense in biological
terms because it is known \cite{Hsieh} \cite{Anderson} that fishing
magnifies fluctuations in exploited species.

To reconstruct the dynamics of $\sigma \left( t\right) $ from the data we
use a standard technique. First using a small time window we compute the
local value of $\sigma \left( t\right) $ by the standard deviation of $%
r\left( t\right) $. For the numerical results presented here a window of $6$
years has been used. Then form the cumulative processes%
\begin{eqnarray}
\sum_{i=1}^{t}\sigma \left( i\right)  &=&\beta _{1}t+R_{1}\left( t\right)  
\notag \\
\sum_{i=1}^{t}\ln \sigma \left( i\right)  &=&\beta _{2}t+R_{2}\left(
t\right)   \label{3.7}
\end{eqnarray}%
$\beta _{1}$ and $\beta _{2}$ being the average values of $\sigma $ and $\ln
\sigma $ and $R_{1}$ and $R_{2}$ the cumulative processes of the
fluctuations about the average. To obtain a model for the fluctuations one
looks for the scaling properties of $R_{1}$ and $R_{2}$, namely the behavior
of $\mathbb{E}\left\vert R_{1}\left( t\right) -R_{1}\left( t+\Delta \right)
\right\vert $ and $\mathbb{E}\left\vert R_{2}\left( t\right) -R_{2}\left(
t+\Delta \right) \right\vert $ as a function of the time lag $\Delta $.

\begin{figure}[p]
\centering
\includegraphics[width=0.8\textwidth]{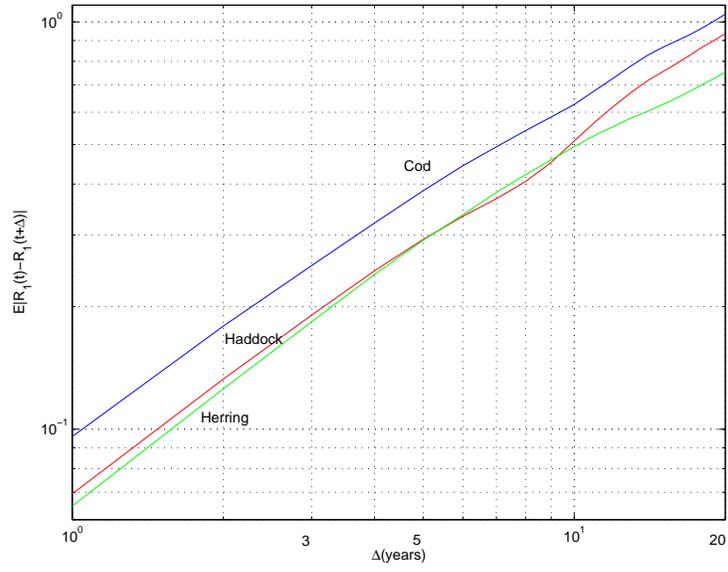}
\caption{The scaling behavior of $R_{1}$}
\end{figure}

\begin{figure}[p]
\centering
\includegraphics[width=0.8\textwidth]{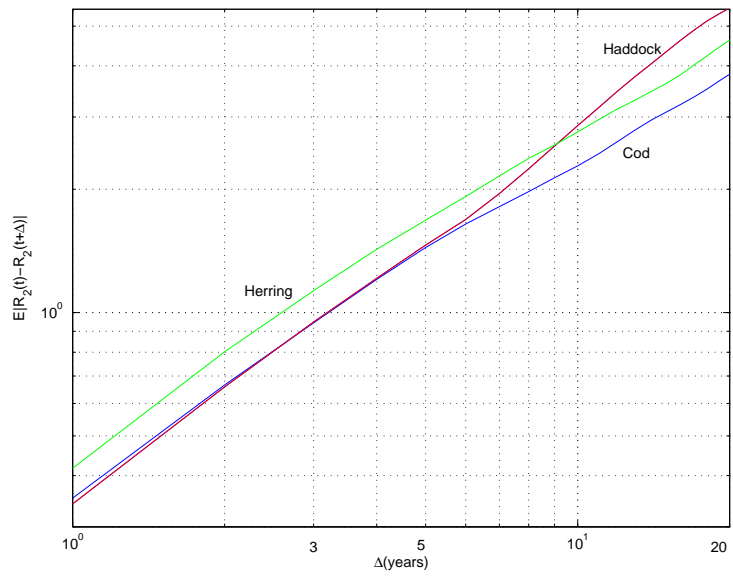}
\caption{The scaling behavior of $R_{2}$}
\end{figure}

The conclusion is that although not perfect, which would not be expected
with data covering at most 68 points, we may assume that $R_{1}$ and $R_{2}$
obey an approximate scaling law with exponents $H$ in the range $0.8-0.9$.
Therefore $R_{1}$ and $R_{2}$ may be modelled by fractional Brownian motion
implying that the fluctuations of $\sigma $ and $\ln \sigma $, away from an
average value, are modeled by Gaussian fractional noise. We have looked for
scaling both for the cumulative $\sigma $ and $\ln \sigma $ to decide which
one would provide a simpler model for the amplitude fluctuations. However,
with the data available, there is no clear decision. Therefore two
alternative models are proposed for the population fluctuations%
\begin{eqnarray}
dS\left( t\right)  &=&\sigma \left( t\right) S\left( t\right) dB_{t}  \notag
\\
\sigma \left( t\right)  &=&\beta _{1}+\alpha _{1}\left( B_{H_{1}}\left(
t\right) -B_{H_{1}}\left( t-1\right) \right)   \label{3.8}
\end{eqnarray}%
or%
\begin{eqnarray}
dS\left( t\right)  &=&\sigma \left( t\right) S\left( t\right) dB_{t}  \notag
\\
\ln \sigma \left( t\right)  &=&\beta _{2}+\alpha _{2}\left( B_{H_{2}}\left(
t\right) -B_{H_{2}}\left( t-1\right) \right)   \label{3.9}
\end{eqnarray}%
From the data the following values are obtained for the Hurst coefficients $%
H_{1}$ and $H_{2}$

\begin{center}
\begin{tabular}{lll}
& $H_{1}$ & $H_{2}$ \\ 
Cod & $0.86$ & $0.87$ \\ 
Haddock & $0.89$ & $0.9$ \\ 
Herring & $0.93$ & $0.87$%
\end{tabular}
\end{center}

From the $H-$values one sees that the dynamics of the fluctuations is a long
range memory process. In addition we have found out that the processes seem
to be species-dependent. For illustration we plot the cumulative amplitude
fluctuations $R_{1}$ $R_{2}$ in the Figs.5 and 6.

\begin{figure}[p]
\centering
\includegraphics[width=0.8\textwidth]{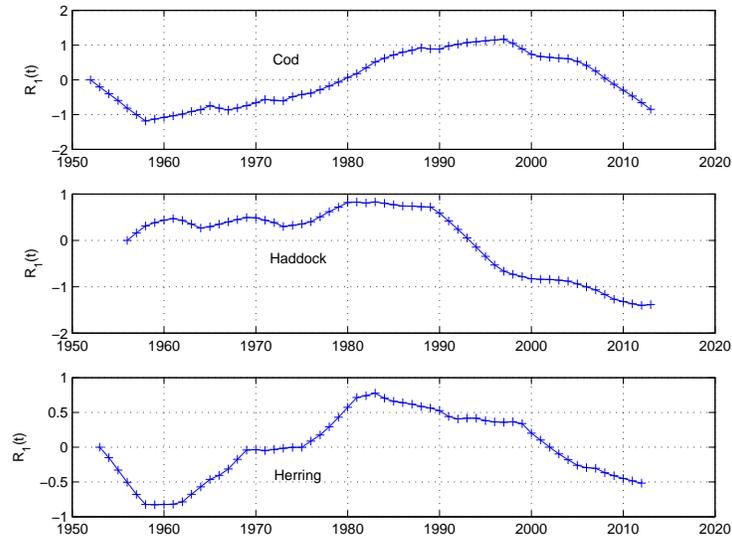}
\caption{Dynamics of the cumulative fluctuations of $\protect\sigma \left(
t\right) $}
\end{figure}

\begin{figure}[p]
\centering
\includegraphics[width=0.8\textwidth]{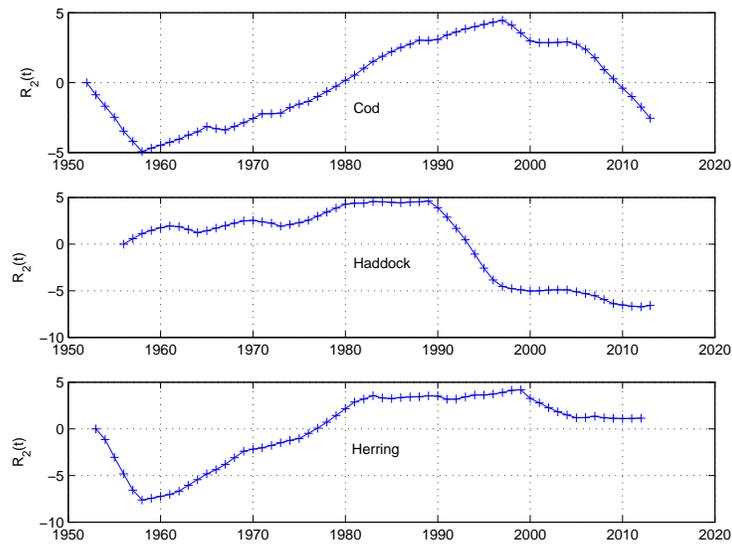}
\caption{Dynamics of the cumulative fluctuations of $\ln \protect\sigma %
\left( t\right) $}
\end{figure}

Processes with such high $H $ values are almost deterministic processes.
This is an important outcome because they may provide clues on the causes of
that particular dynamics.

In conclusion, this methodology of separation of the pure random features
from the non-trivial dynamics of amplitude fluctuations may be useful for
the analysis of other natural processes. In fact it has a solid mathematical
basis as a consequence of the stochastic analysis result mentioned in (\ref%
{2.13}). The dynamics of exploited fish populations provides a good example
of a phenomenon where the long range dependence features appear not at the
level of second order correlations but only through the analysis of higher
order effects.

\section{Discussion}

1 - Our paper was motivated by the concerns raised in Niwa \cite{Niwa} that
population variability corresponds to a geometric random walk and,
consequently, the exploited population trajectory is a series of of random
uncorrelated abundances over time. The annual change of population abundance
is the result of many \textquotedblleft shocks\textquotedblright , including
recruitment variability, natural mortality (e.g. predation and competition)
and varying fishing pressure. Despite these collection of independent
drivers our results suggests that each fish stock dynamics exhibit long
range dependence. Moreover, from the empirically found $H$-values associated
to the fluctuations, one sees that the long range memory process seems to be
species dependent. This is the reason leading to Niwa's random walk
conclusion because, when averaging over species, the $\sigma \left( t\right) 
$ process in Eqs. (\ref{3.8}) and (\ref{3.9}) would be simply replaced by a
fixed number $\overline{\sigma \left( t\right) }$ (the average of $\sigma
\left( t\right) $).

2 - A full discussion of whatever is actually determining the stochastic
process for each species is beyond the scope of this study but the question
why fish populations fluctuate has generated much attention from fishery
scientists and marine ecologists over the past century. Three general
hypotheses have been proposed to answer this question: (i) species
interactions generate fluctuating and cyclic population dynamics; (ii)
nonlinearity in single-species dynamics generates deterministic
fluctuations; and (iii) changes in the environment determines variation in
vital rates and recruitment, which in turn drive variation in abundance \cite%
{Shelton} \cite{Anderson} \cite{Overland}. These are not mutually exclusive
hypotheses as all three could act together to increase variability. For
exploited species, fishing can also vary from year to year and translate
directly into population variability or could interact with the other
drivers to enhance fluctuations in fish abundance \cite{Anderson} \cite%
{Turchin}.

Long range trends are frequently related to external forcing on the
populations and are usually derived from human exploitation or environmental
change. The dependence in the population growth observed for the haddock,
herring and cod stocks could be derived from the different and varying
exploitation regimes and/or from large-scale environmental changes as the
North Atlantic Oscillation, that can induce low (or high) productivity
regimes in fish recruitment (e.g. \cite{Leif} \cite{Borges} \cite{Brander}).

Bj\o rnstad et al. \cite{Bjornstad} have shown that short-term variability
in recruitment caused by environmental change, combined with intercohort
interactions can be echoed through the population age structure inducing
persistent cycles and long-term fluctuations. It is also known \cite%
{Fromentin2} that noise in recruitment combined with a large number of
classes of spawners could lead to long-term variations in spawning stock
biomass and yields, as well as to regular cycles, depending on the lifespan
of the species. Furthermore, the stocks analyzed here showed species
specific stochastic processes that should be analyzed considering the
contrasting life history traits. The distinct growth rates, age at maturity,
spawning duration and lifespan are characteristics that make some fish
stocks more or less vulnerable to exploitation and environmental conditions 
\cite{Anderson} \cite{Reynolds}. The small bodied and younger herring
population should be less able to smooth out environmental fluctuations and
more prone to exhibit unstable dynamics due to changing demographic
parameters.

3 - There is no doubt that fisheries management would profit from a clearer
understanding of the mechanisms determining the dynamics of the amplitude of
the fluctuations in exploited fish stocks. The failure of many fish stocks,
despite the implemented management measures, to recover rapidly to former
levels of abundance, might arguably be related to the long range memory
observed in this study. However, acknowledging the uncertainty arising from
the reduced number of species analyzed stresses the importance of retaining
more time series of population abundance over long periods, if the aim is to
detect and describe the underlying mechanisms that drive population
variability.


\begin{thebibliography}{99}
\bibitem{Samorod} G. Samorodnitsky; \textit{Long Range Dependence},
Foundations and Trends in Stochastic Systems 1 (2006) 163--257.

\bibitem{Dominique} G. Dominique; \textit{How can we define the concept of
long memory? An econometric survey}, Econometric reviews 24 (2005) 113--149.

\bibitem{TaqquBook} P. Doukhan, G. Oppenheim, and M. S. Taqqu; \textit{%
Theory and Applications of Long-Range Dependence, }Springer, Berlin 2003.

\bibitem{SamoTaqqu} G. Samorodnitsky and M. S. Taqqu; \textit{Stable
non-Gaussian processes: Stochastic models with infinite variance}, Chapman
and Hall, New York 1994.

\bibitem{Nualart} D. Nualart; \textit{The Malliavin calculus and related
topics}, Springer, Berlin 2006.

\bibitem{Granger} C. Granger and R. Joyeux; \textit{An introduction to
long-memory time series and fractional differencing}, J. of Time Series
Analysis 1 (1980) 15-30.

\bibitem{Hosking} J. Hosking;\textit{\ Fractional differencing}, Biometrika
68 (1981) 165-176.

\bibitem{Bradley} R. Bradley; \textit{Basic properties of strong mixing
conditions. A survey and some open questions}, Probability Surveys 2 (2005)
107-144.

\bibitem{Dickson} R. Dickson and K. Brander; \textit{Effects of a changing
windfield on cod stocks of the North Atlantic}, Fisheries Oceanography 2
(1993)124-153.

\bibitem{Bjornstad} O. Bj\o rnstad, J. M. Fromentin, N. C. Stenseth and J. Gj%
\o s\ae ter; \textit{Cycles and trends in cod populations}, Proceedings of
the National Academy of Sciences USA, 96 (1999) 5066--5071.

\bibitem{Fromentin} J. M. Fromentin, R. M. Myers, O. Bj\o rnstad, N. C.
Stenseth, J. Gj\o s\ae ter and H. Christie; \textit{Effects of
density-dependent and stochastic processes on the stabilization of cod
populations}, Ecology 82 (2001) 567--579.

\bibitem{Niwa} H.-S. Niwa; \textit{Random-walk dynamics of exploited fish
populations}, ICES Journal of Marine Science, 64 (2007) 496--502.

\bibitem{ICES} %
http://www.ices.dk/community/advisory-process/Pages/Latest-Advice.aspx

\bibitem{Hsieh} C.-h. Hsieh, C. S. Reiss, J. R. Hunter, J. R. Beddington, R.
M. May and G. Sugihara; \textit{Fishing elevates variability in the
abundance of exploited species}, Nature 443 (2006) 859-862.

\bibitem{Anderson} C. N. K. Anderson, C.-h. Hsieh, S. A. Sandin, R. Hewitt,
A. Hollowed, J. Beddington, R. S. May and G. Sugihara; \textit{Why fishing
magnifies fluctuations in fish abundance}, Nature 452 (2008) 835-839.

\bibitem{Leif} C. S. Leif, G. Ottersen, K. Brander, K. Chan and N. C.
Stenseth; \textit{Cod and climate: effect of the North Atlantic Oscillation
on recruitment in the North Atlantic}, Marine Ecology Progress Series 325
(2006) 227--241.

\bibitem{Shelton} A. O. Shelton and M. Mangel; \textit{Fluctuations of fish
populations and the magnifying effects of fishing}. Proceedings of the
National Academy of Sciences USA, 108 (2011)7075--7080.

\bibitem{Turchin} P. Turchin and A. D. Taylor; \textit{Complex dynamics in
ecological time series}. Ecology 73 (1992) 289--305.

\bibitem{Reynolds} J. Reynolds, S. Jennings and N. K. Dulvy; \textit{Life
histories of fishes and population responses to exploitation}, In :
Conservation of Exploited Species pp. 147-168, J. D. Reynolds, G. M. Mace,
K. H. Redford and J. G. Robinson (eds.), Cambridge University Press,
Cambridge 2001.

\bibitem{Overland} J. E. Overland, J. Alheit, A. Bakun, J. W. Hurrell, D. L.
Mackas and A. J. Miller; \textit{Climate controls on marine ecosystems and
fish populations}, Journal of Marine Systems 79 (2010) 305--315.

\bibitem{Fromentin2} J. Fromentin and A. Fonteneau; \textit{Fishing effects
and life history traits: a case study comparing tropical versus temperate
tunas}. Fisheries Research 53 (2001) 133-150.

\bibitem{Borges} M. F. Borges, H. C. Mendes and A. M. P. Santos; \textit{%
Sardine (Sardina pilchardus) recruitment is strongly affected by climate
even at high spawning biomass in West Iberia/Canary upwelling system} in
Science and Management of Small Pelagics, S. Garcia, M. Tandstad and A. M.
Caramelo (eds.), FAO Fisheries and Aquaculture Proceedings 18 (2011) 237-244.

\bibitem{Brander} K. Brander; \textit{Cod recruitment is strongly affected
by climate when stock biomass is low}, ICES Journal of Marine Science 62
(2005) 339-343.
\end{thebibliography}
\end{document}